\documentclass[aps,preprint,superscriptaddress]{revtex4-2}
\usepackage{graphicx}
\usepackage{dcolumn}
\usepackage{bm}
\usepackage[usenames]{color}
\usepackage{amsmath}
\usepackage{amssymb}
\usepackage{latexsym}

\renewcommand{\thefigure}{\arabic{figure}}

\newcommand{\CRO}{Ca$_2$RuO$_4$}
\newcommand{\SRO}{Sr$_2$RuO$_4$}

\begin{document}

\title{
Bose glass in \CRO~nanofilms
}

\author{Hiroyoshi Nobukane}
\affiliation{Department of Physics, Hokkaido University, Sapporo, 060-0810, Japan}
\affiliation{Center of Education and Research for Topological Science and Technology, Hokkaido University, Sapporo, 060-8628, Japan}

\author{Koki Hirose}
\affiliation{Department of Physics, Hokkaido University, Sapporo, 060-0810, Japan}

\author{Kakeru Isono}
\affiliation{Department of Physics, Hokkaido University, Sapporo, 060-0810, Japan}

\author{Mizuki Higashiizumi}
\affiliation{Department of Applied Physics, Hokkaido University, Sapporo, 060-8628, Japan}

\author{Masahito Sakoda}
\affiliation{Department of Applied Physics, Hokkaido University, Sapporo, 060-8628, Japan}

\author{Korekiyo Takahashi}
\affiliation{Department of Applied Physics, Hokkaido University, Sapporo, 060-8628, Japan}

\author{Satoshi Tanda}
\affiliation{Center of Education and Research for Topological Science and Technology, Hokkaido University, Sapporo, 060-8628, Japan}
\affiliation{Department of Applied Physics, Hokkaido University, Sapporo, 060-8628, Japan}

\begin{abstract}
    \textbf{
    Weak localization of bosons can give rise to an exotic quantum phase known as a Bose glass, characterized by the absence of global phase coherence yet finite conductivity. 
    This phase is crucial in understanding the interplay between disorder, interactions, and superconductivity, especially in two-dimensional and strongly correlated systems.
    Here we report the presence of the Bose glass phase in the weak localization region of ruthenium oxide \CRO.
    The electrical resistance exhibits a characteristic logarithmic temperature dependence in this phase, $\rho \sim \ln(1/T)$.
    Through $\beta$-function scaling analysis, we observed “vertical flow,” indicating unconventional scaling behavior associated with localized bosonic states. Our results suggest the existence of bosons—Cooper pairs—persisting up to high temperatures around 220 K and that these bosons undergo weak localization.
    In the Bose glass phase, vortices are found to have a dual relationship with the localized Cooper pairs, enabling their motion and resulting in finite resistance despite the presence of bosonic order. We identified two quantum critical points: one between the Bose glass and superconducting phases and another between the Bose glass and Mott insulating phases, allowing us to extract the corresponding quantum sheet resistances.
    We revealed that the ground state of the \CRO~ changes depending on the localization strength.
    Thinning the \CRO~ corresponds to controlling the electronic correlation by relieving the distortion in RuO$_6$ octahedra.
    These findings offer significant insights into the interplay between electronic correlations and bosonic transport, with important implications for studying high-temperature superconductors based on perovskite structures.
    }
\end{abstract}

\maketitle

\noindent
\subsection*{Introduction}
A superconductor-insulator transition (SIT) occurs in two dimensions (2D) by tuning the film thickness~\cite{haviland1989onset,tanda1992bose}, magnetic field~\cite{paalanen1992low}, electric field~\cite{bollinger2011superconductor}, and Josephson junction arrays~\cite{fazio2001quantum}, a quantum phase transition. 
One of the characteristics of a superconducting state is zero resistance. 
This property is because the bosonic Cooper pairs conduct without dissipation. 
The characteristics of the boson also appear in the finite resistance of the 2D SIT when the external parameters are controlled in the vicinity of $T = 0$. 
In the superconducting phase of 2D SIT, the Cooper pairs condensate, and the quantized vortex localizes. 
On the other hand, in the insulating phase, the Cooper pairs localize, and the vortex flow leads to the finite resistance. 
Cooper pairs and vortices have a self-dual relationship, and the universal quantum critical sheet resistance $R_{\Box / \mathrm{layer}}$ is expressed as $R_{\Box / \mathrm{layer}}=h/4e^2 = 6.45~\mathrm{k}\Omega$, where $h$ is Planck's constant and $e$ is the electric charge.
The superinsulating state has been reported in titanium nitride films~\cite{vinokur2008superinsulator} and indium oxide~\cite{sambandamurthy2005experimental}, which is dual to the superconducting state.
Numerous reports have been on research regarding intermediate boson metals~\cite{das1999existence,kapitulnik2019colloquium}.
Chern-Simons theory has predicted that an anyonic topological ground state can be derived using Cooper pairs and vortices in two-dimensional boson systems\cite{diamantini2020bosonic}.

Electric transport properties can detect the behavior of localized Cooper pairs.
Takahashi and co-workers have recently reported the weak localization of Cooper pairs (Bose glass) and its transition to Fermi glass in two-dimensional superconductors of copper oxides and Pb thin films~\cite{takahashi2023bose}.
They performed a scaling analysis in the weak localization regime. 
They clarified the behavior of the Bose glass phase (weak localization of Cooper pairs) as a "vertical flow" from the flow of the $\beta$-function, in addition to the Fermi glass discussed so far~\cite{tanda1992bose,kagawa1996superconductor}.
In Fermi glass, the temperature dependence of conductivity shows the $\sigma \sim \mathrm{ln}~T$ characteristic due to Anderson localization. In contrast, in Bose glass, the temperature dependence of resistivity follows $\rho \sim \mathrm{ln}~(1/T)$~\cite{takahashi2023bose,das1998weakly}.
These analysis methods are decisive for determining whether the localized state is a boson or a fermion.
It is important whether weak localization is caused by a disorder or strongly correlated electron systems~\cite{byczuk2010anderson}.
It is not apparent whether the weak localization theory and the $\beta$-function scaling theory can be applied to the strongly correlated electron systems.
Therefore, we focused on the SIT in nanoscale ruthenium oxide thin films to investigate the relationship between disorder and interacting electron systems.

Layered ruthenate oxide \CRO~ exhibits a variety of ground states, from superconductivity to a ferromagnetic metal and a Mott insulator, depending on chemical substitution~\cite{nakatsuji2004mechanism,cao2000ground}, physical pressure~\cite{nakamura2002mott}, photo-doping~\cite{li2025time}, electric field~\cite{nakamura2013electric}, and bias-current~\cite{Nobukane_CRO}.
The release of three types of distortions -tilting, rotating, and flattening- in the RuO$_6$ octahedra plays an important role in the significant changes in electronic properties.
\SRO, in which Ca is replaced with Sr, exhibits unconventional superconductivity at low temperatures~\cite{maeno2024thirty}. 
Our recent research~\cite{Nobukane_CRO} reported that high-temperature superconductivity with the onset temperature $T_c^{\mathrm{onset}}$ of 100~K appears when the film thickness is reduced to the nanometer range.
In Ref.~\cite{Nobukane_CRO}, it is notable that the resistance of the superconducting sample also increases gradually at temperatures above the $T_c^{\mathrm{onset}}$.
Thus, we have investigated the localization of bosons in \CRO.
There have been reports of theoretical studies into the high-temperature superconductivity of ruthenates~\cite{carrasco2025orbital}.
Research into nickel oxide thin film superconductivity is also progressing~\cite{li2019superconductivity,zhang2024high,zhou2025ambient}.
The superconductivity in nickelates resembles that of high-temperature superconductors in nanoscale \CRO~ thin films~\cite{Nobukane_CRO}.

In this paper, we found a Bose glass phase between the superconducting and Mott insulating phases in a quantum phase transition.
We determined the quantum sheet resistances at two critical points.
We observed a transition from Bose glass to Fermi glass in the weakly localized region at high temperatures.
Surprisingly, the Bose glass phase was at the highest temperature of about 220~K.
This means that Cooper pairs are localized in the Bose glass phase and that vortices are driven.
The electrical resistivity at room temperature can determine the \CRO~ ground state at low temperatures.
Maintaining the weak localized state to low temperatures makes it possible to induce the Bose glass and superconducting phases.
We have succeeded in controlling electron correlation by changing the film thickness.

\noindent
\subsection*{Results and discussion}

\begin{figure}[t]
\begin{center}
\includegraphics[width=1.0\linewidth]{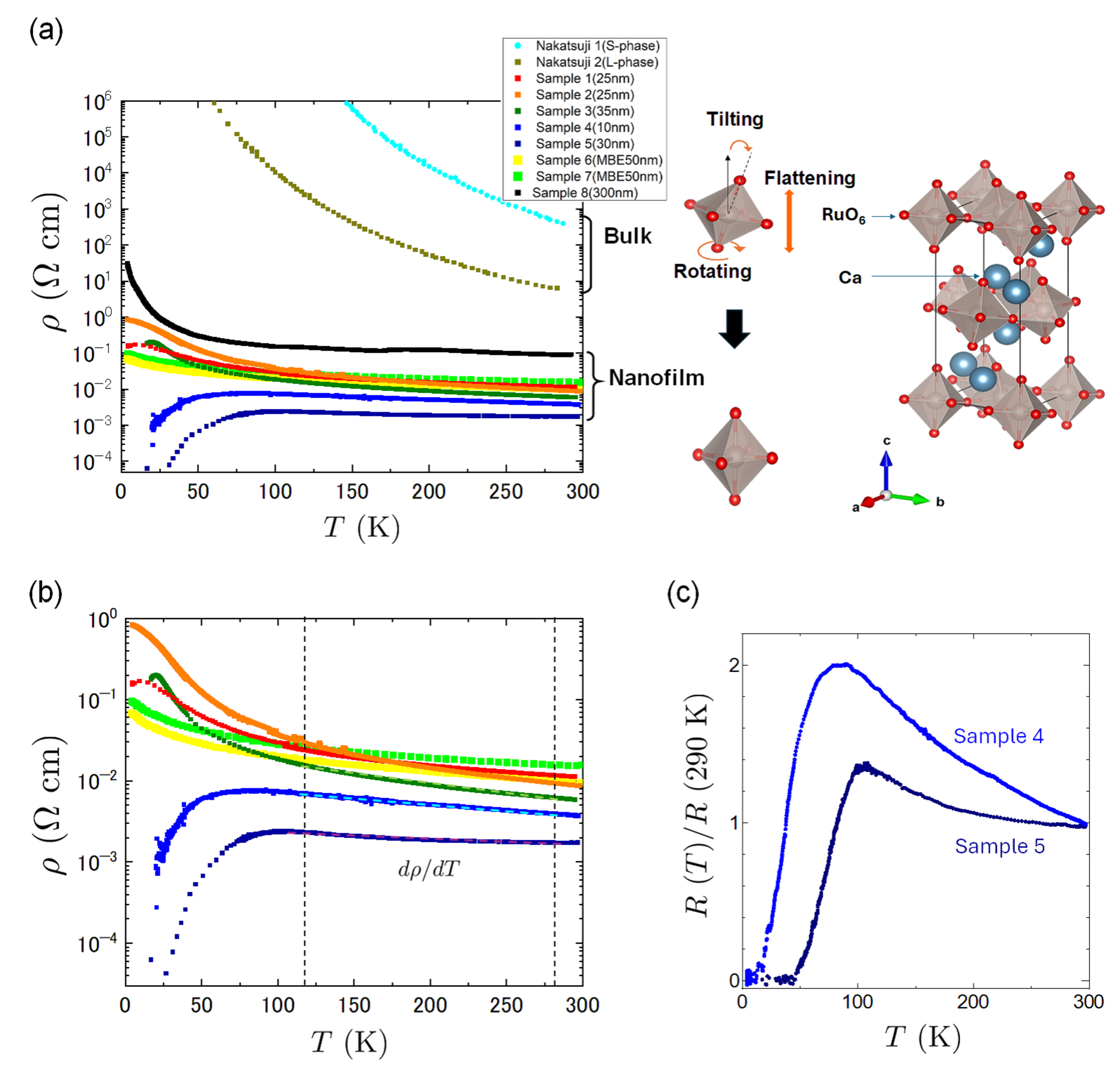}
\caption{
    \textbf{Thickness-tuned superconductor-insulator transition in \CRO.}
    (a) Temperature dependence of resistivity in nanoscale thin film single crystals (samples 1-5 and 8), MBE thin films (samples 6 and 7), and bulk single crystals (Nakatsuji 1 and 2).
    The S phase and L phase data in Ref~\cite{nakatsuji1997} are cited as Nakatsuji 1 and Nakatsuji 2, respectively.
    By reducing the thickness of the sample to the nanoscale, the resistivity decreases.
    The right panel shows the crystal structure of \CRO~ and the elimination of distortions on the RuO$_6$ octahedra in the nanoscale thin film.
    (b) Enlargement of the dependence of resistivity on temperature in samples 1-7.
    The increase in resistivity is expressed as $d\rho (= \rho(120~\mathrm{K}) - \rho(280~\mathrm{K}))/dT$.
    (c) Temperature dependence of the normalized resistance in samples 4 and 5. 
    High-temperature superconductivity has been achieved in thin films at ambient pressure.
}
\label{figure1}
\end{center}
\end{figure}

Figure~\ref{figure1} shows the temperature dependence of the resistance for different film thicknesses of Ca$_2$RuO$_4$.
The bulk of Ca$_2$RuO$_4$ shows a structural transition of 354~K at ambient pressure. 
Below the structural transition temperature, it shows the behavior of variable range hopping (VRH), the Mott insulator with strong electron correlation~\cite{nakatsuji1997}. 
In Fig.~\ref{figure1}(a), the results of the L and S phases of the Ca$_2$RuO$_4$ in Ref.~\cite{nakatsuji1997} are plotted.
The L and S phases are characterized by longer or shorter $c$-axis parameters.
The L phase has less distortion and lower resistivity than the S phase.
On the other hand, the resistivity of our thin film samples at room temperature is 3 to 4 orders of magnitude smaller than that of the bulk sample.
In general, thin film samples exhibit higher resistivity due to the more significant contribution of impurities. 
Our results show a surprising difference from what is typically expected.
This is because forming a thin film in the layered ruthenates releases the distortion in RuO$_6$ octahedra.
The distortion in the RuO$_6$ octahedra, rather than impurities, affects the electronic correlation.

\begin{figure}[t]
\begin{center}
\includegraphics[width=1\linewidth]{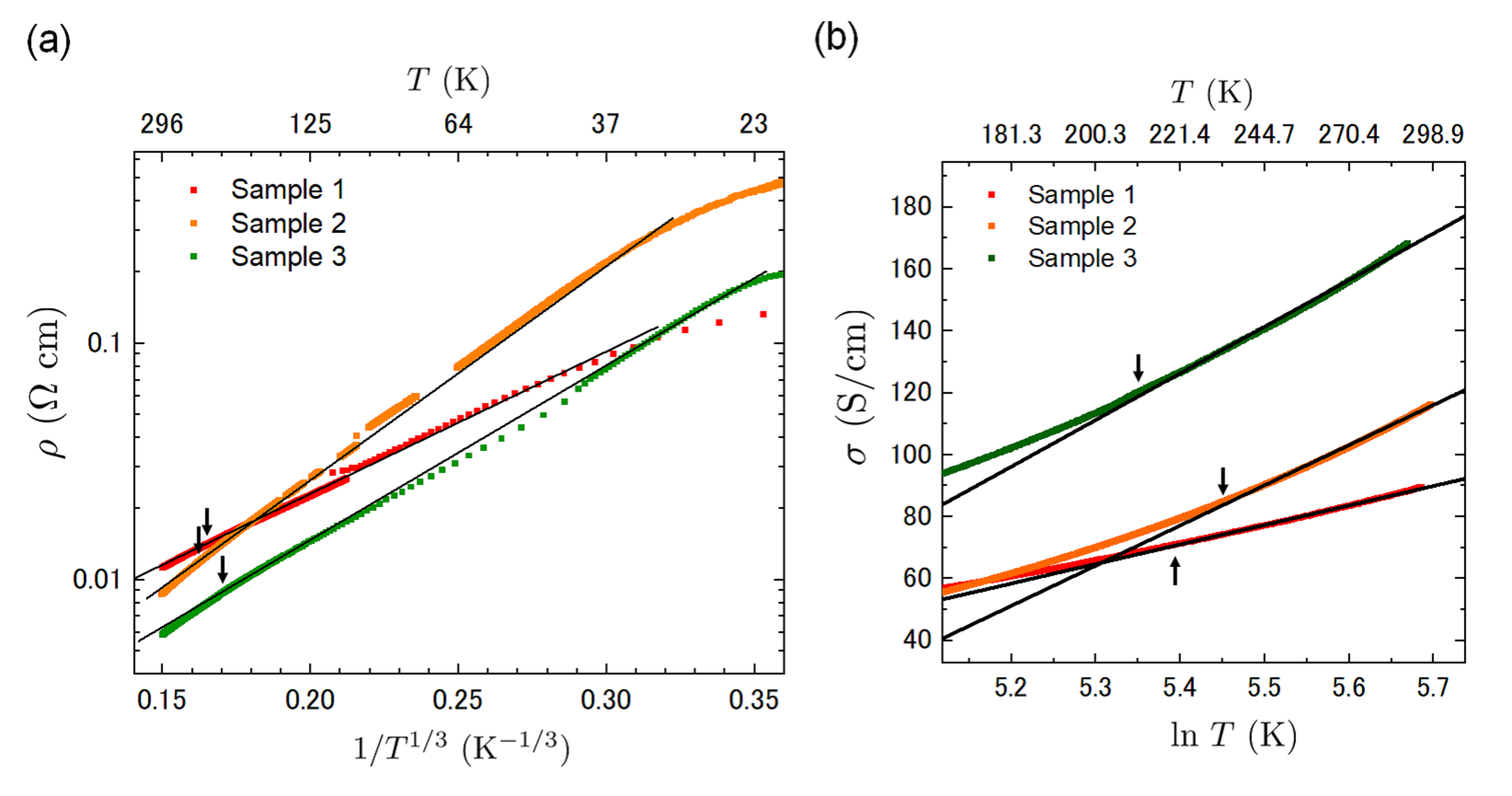}
\caption{
    \textbf{Transition from Fermi glass to Mott insulator with decreasing temperature.}
     (a) Resistivity as a function of $T^{-1/3}$ in samples 1-3. The resistivity fits a linear black line, suggesting 2D VRH conduction. 
     The black arrows indicate the temperature at which the linear fit of VRH conduction deviates.
     (b) In the high-temperature region, the data fit well to the function of $\sigma \sim \mathrm{ln}~T$ as shown in black lines (samples 1-3).
}
\label{figure2}
\end{center}
\end{figure}

In thin film samples, the resistance gradually increases as temperature decreases from room temperature.
We first discuss the results of samples 1-3 in Fig.~\ref{figure2}.
As with the L and S phases of the bulk \CRO~sample, samples 1-3 resistance increases at low temperatures.
A two-dimensional variable range hopping conduction $R_\square(T)=R_0 \exp (T_0/T)^{1/3}$ fits this well, indicating the behavior of the Mott insulator in Fig.~\ref{figure2}(a).
On the other hand, the increase in resistance from room temperature to 224~K for sample 1, 233~K for sample 2, and 208~K for sample 3, respectively, shows $\sigma \sim \mathrm{ln}~T$ dependence as shown in Fig.~\ref{figure2}(b).
In two dimensions, it is theoretically proposed that the weak localization of fermions behaves as $\sigma \sim \mathrm{ln}~T$ and that of bosons behaves as $\rho \sim \mathrm{ln}~(1/T)$~\cite{das1998weakly}.
Therefore, the behavior from room temperature to around 220~K indicates that the conduction electrons are in a weakly localized Fermi glass.
The results show that samples 1-3 transition from the Fermi glass to the Mott insulator near 220~K.
The transition temperature of bulk crystals to the Mott insulator is 354 K, but in thin films, it shifts to a lower temperature.

\begin{figure}[t]
\begin{center}
\includegraphics[width=1\linewidth]{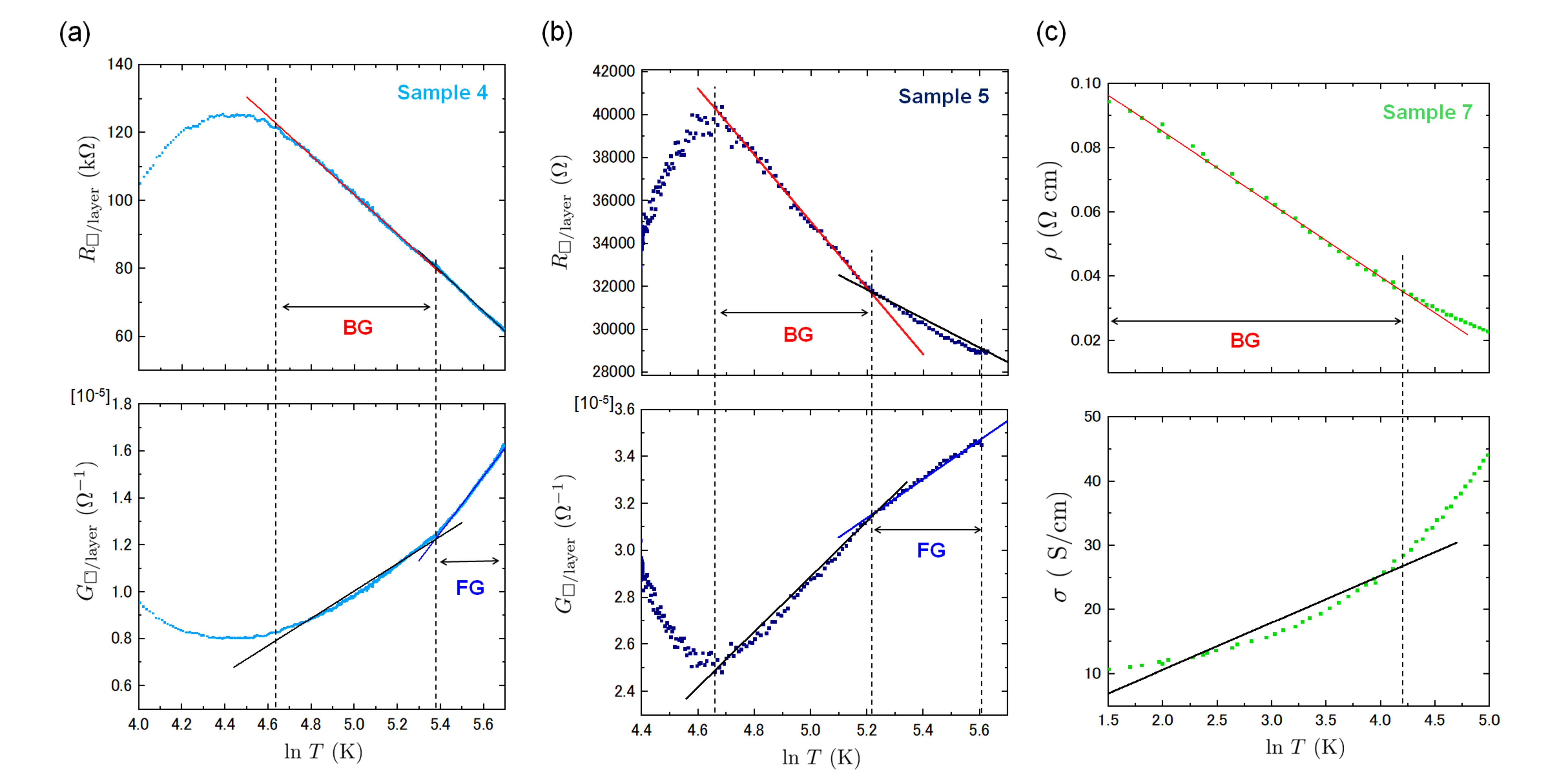}
\caption{
    \textbf{Transition from Fermi glass to Bose glass.}
    (a) and (b) $\mathrm{ln}~T$ dependence of sheet resistance $R_{\Box / \mathrm{layer}}$ and sheet conductivity $G_{\Box / \mathrm{layer}}$ for samples 4 and 5.
    For sample 4, the data aligns well with the $R_{\Box / \mathrm{layer}}\sim \mathrm{ln}~(1/T)$ fitting than $G_{\Box / \mathrm{layer}}\sim \mathrm{ln}~T$ in the $101<T<220$~K range. The data for sample 5 also fits well with the $R_{\Box / \mathrm{layer}}\sim \mathrm{ln}~T$ fitting in the $105<T<183$~K range.
    (c) $\mathrm{ln}~T$ dependence of resistivity and conductivity for MBE thin films.
}
\label{figure3}
\end{center}
\end{figure}

Samples 4 and 5 show 2D superconducting behavior below $T_c^{\mathrm{onset}}$.
Figure~\ref{figure1}(d) shows the result.
In this paper, we analyze the weak localization of bosons in terms of the resistance increase from room temperature to the $T_c^{\mathrm{onset}}$.
Figures~\ref{figure3} (a) and (b) plot ln~$T$ dependence of sheet resistance $R_{\Box / \mathrm{layer}}$ and sheet conductivity $G_{\Box / \mathrm{layer}}$ for samples 4 and 5, respectively.
For sample 4 (Fig.~\ref{figure3}(a)), our data are in better agreement with the $R \sim \mathrm{ln}~(1/T)$ fitting than with $G \sim \mathrm{ln}~T$ in the range $101 < T < 220$ K. 
Sample 5 also shows a good $R \sim \mathrm{ln}~(1/T)$ fit in the temperature range $105 < T < 183$ K.
From around 220~K to room temperature, the samples show Fermi glass behavior.
In samples 4 and 5, we observed the transition from weakly-localized fermi glass to Bose glass.
Interestingly, we observed the behavior of the Bose glass from around 220~K, which is higher than the $T_c^{\mathrm{onset}}$.
Our results indicate the appearance of localized Cooper pairs and moving vortices at a high temperature of around 220 K for sample 4 and 183~K for sample 5.
Notably, when the phases of the Cooper pairs are aligned, a significant decrease in electrical resistance is observed.

Let us discuss the behavior of the Bose glass in magnetic properties. 
In the magnetization measurement of nanoscale samples, after the ferromagnetic transition at 180~K~\cite{Nobukane_CRO}, the diamagnetism occurs from around 140~K, which is close to the temperature of the Fermi glass-Bose glass transition that observed in the electrical resistance.
In bulk, \CRO~, an applied DC induced the diamagnetism from around 220~K~\cite{sow2017retracted,mattoni2024challenges}.
It has been concluded that the behavior was due to the local Joule heating effect, and Ref.~\cite{sow2017retracted} has already been retracted.
However, observing diamagnetic components from around 220~K and the Bose glass transition temperature occurs in the same range.
This diamagnetism at high temperatures may be related to the localization of Cooper pairs.

Interestingly,  MBE thin film samples 6 and 7 show the behavior of Bose glass below 67 and 81~K, respectively, in Fig.~\ref{figure3}(c), but do not show zero resistance as shown in Fig.~\ref{figure1}(b) (see Supplementaty Fig.~\ref{Supplfigure1}).
The behavior of the MBE samples is near the quantum critical point of the SIT.
This is because a superconducting path has not been formed from one sample end to the other.
The length $c=11.592$~\r{A} of the $c$-axis of MBE films is shorter than the 12.278~\r{A} of nanoscale single crystals.
Thus, we think that the distortion of the RuO$_6$ octahedra has a significant effect.
We have clarified that Bose glass appears in thin film samples at high temperatures. 
When substantial distortion in RuO$_6$ octahedra exists in thin film samples, the Mott transition (to strong localization) occurs around 220 K. 
When the distortion is weak, bosons are produced and exhibit weak localization.

\begin{figure}[t]
\begin{center}
\includegraphics[width=1\linewidth]{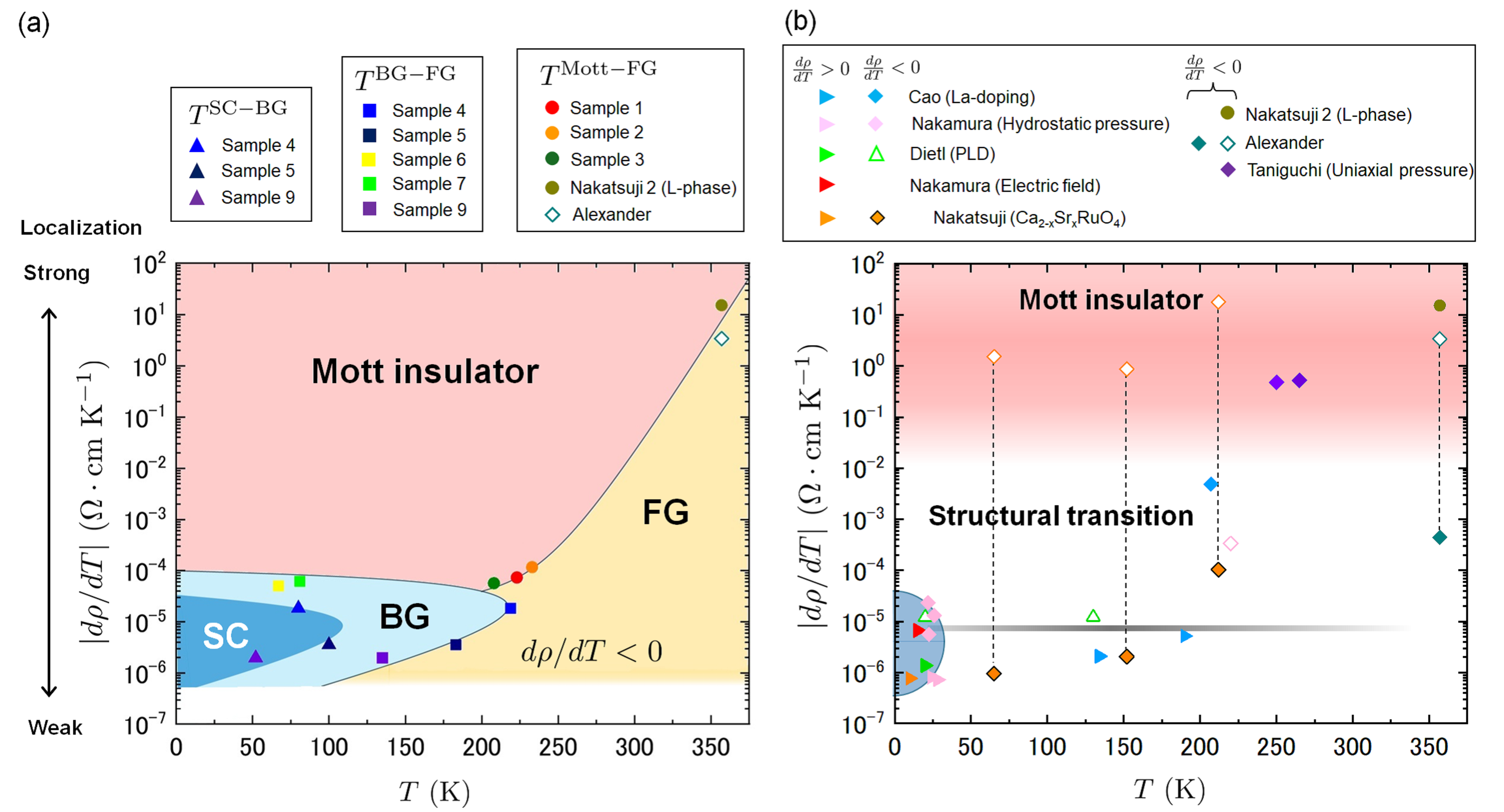}
\caption{
    \textbf{Appearance of the Bose glass phase.}
    (a) The phase diagram in \CRO~nanofilms.
    $d\rho/dT$ represents the strength of localization.
    The distortion on RuO$_6$ octahedra is significant in thick films, and a strong localization (Mott transition) occurs.
    On the other hand, in nanoscale thin film samples, the distortion in RuO$_6$ octahedra is released so that it becomes weakly localized, and the Bose glass phase and superconducting phase appear.
    The squares represent the transition temperature from the Fermi glass (FG) to the Bose glass (BG) phase, and the triangles show the transition temperature from the BG to the superconducting (SC) phase.
    The circles exhibit the transition from the FG to the Mott insulating phase.
    The two points in the top right are plotted using the data reported in the bulk \CRO~\cite{alexander1999destruction,nakatsuji1997}.
    (b) The reported results~\cite{cao2000ground,alexander1999destruction,nakatsuji1997,nakatsuji2004mechanism,dietl2018tailoring,taniguchi2013anisotropic,nakamura2002mott,nakamura2013electric} for the \CRO~ are shown as $d\rho/dT$ vs. the Mott transition or ferromagnetic metal transition temperature.
    The vertical black dotted line shows $d\rho/dT$ before and after the structural transition.
}
\label{figure4}
\end{center}
\end{figure}

Thinning \CRO~films is more about controlling electron correlation than dealing with impurity-induced disorder effects.
Here, we define the increase in resistivity from around room temperature as $d\rho (= \rho(120~\mathrm{K}) - \rho(280~\mathrm{K}))/dT$ to express the strength of localization as shown in Fig.~\ref{figure1}(b).
Figure~\ref{figure4}(a) plots $d\rho/dT$ vs. transition temperature for each sample, along with the Mott-Fermi glass transition temperature ($\bigcirc$), the Bose glass-Fermi glass transition temperature ($\square$), and the superconductor-Bose glass transition temperature ($\triangle$).
For comparison, Figure~\ref{figure4}(b) shows the relationship between $d\rho/dT$ and the Mott transition or ferromagnetic metal transition temperature, based on the results of the CRO reported so far~\cite{cao2000ground,alexander1999destruction,nakatsuji1997,nakatsuji2004mechanism,dietl2018tailoring,taniguchi2013anisotropic,nakamura2002mott,nakamura2013electric}.
We found the Bose glass and superconducting phases in nanoscale thin films with reduced distortion of RuO$_6$.
The smaller the $d\rho/dT$ (the weaker the localization), the more likely the ground state becomes in the Bose glass or superconducting phase.
High-temperature superconductivity appears through the mediation of the Bose glass phase.
In bulk \CRO~(Fig.~\ref{figure4}(b)), the Bose glass and superconducting phases are suppressed.
By reducing the thickness of the film, the distortion in RuO$_6$ octahedra is eliminated. 
Reducing the distortion in RuO$_6$ octahedra leads to a decrease in localization.
Manipulating electron correlation through thin films has significantly reduced electrical resistivity.
We note that the effect of impurity-induced disorder in thin films cannot explain this phenomenon.

In bulk \CRO~crystals, a ferromagnetic metal transition occurs at approximately 20 K, as indicated by the decrease in resistance (the blue region in Fig.~\ref{figure4}(b)).
In the Supplementary Fig.~\ref{Supplfigure2}, we analyzed a superconducting fluctuation using the Aslamazov-Larkin fit in two dimensions~\cite{aslamasov1968influence} on the 2.6 GPa data reported in Ref.~\cite{nakamura2013electric}. 
Superconductivity coexisting with ferromagnetism may also occur in bulk crystals.

\begin{figure}[t]
\begin{center}
\includegraphics[width=1\linewidth]{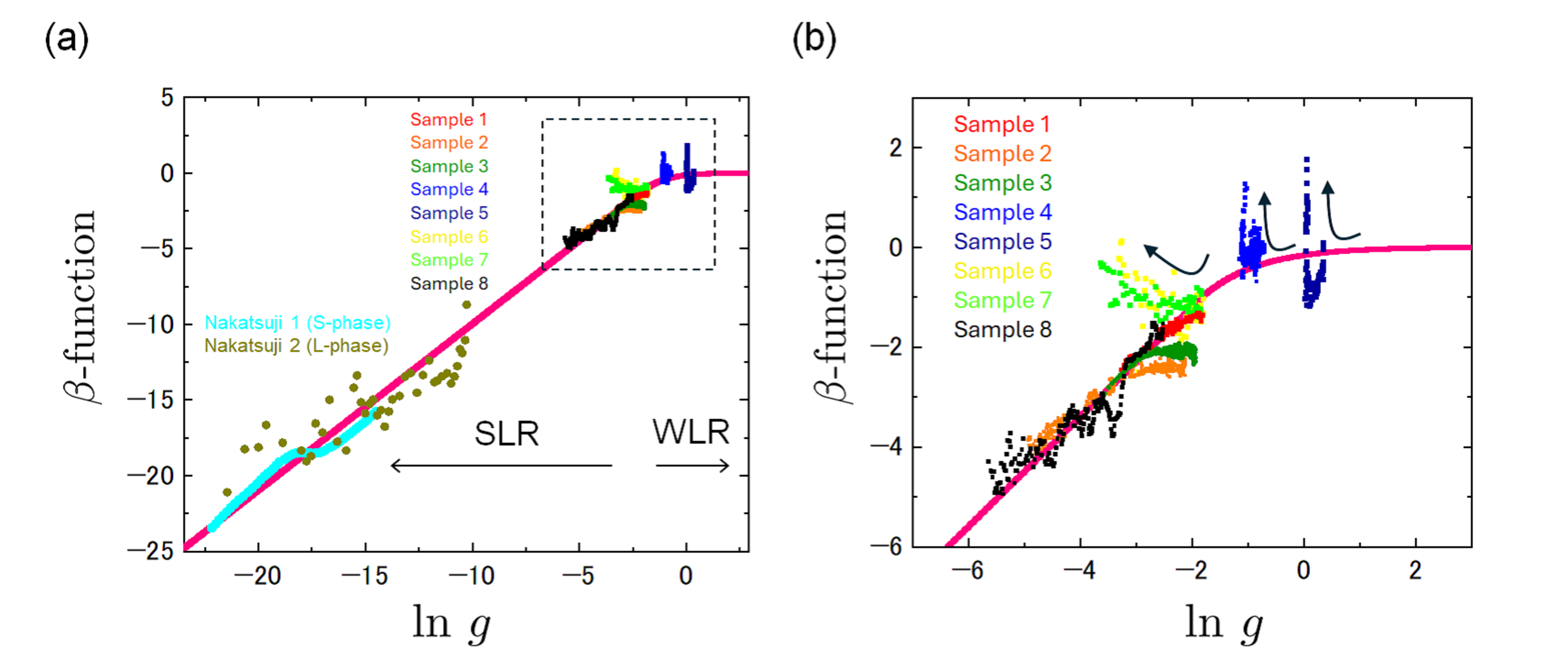}
\caption{
    \textbf{Behavior of scaling of dimensionless conductance in the \CRO.}
    (a) The experimental data show that the conductance of all samples fits a single universal curve. 
    The solid pink curve shows Vollhardt and W\"{o}lfle $\beta_{\mathrm{VW}}$-function. 
    Our samples indicate the $-1/g$ dependence in the weak localized region and $\mathrm{ln}~g$ dependence in the strong localized region.  
    (b) The dotted square in (a) is enlarged. 
    The weak fermion localization behaves along the $\beta_{\mathrm{VW}}(g)$-function.
    On the other hand, samples 4-7 show a flow perpendicular to the line of $\beta_{\mathrm{VW}}(g)$. 
    This perpendicular flow is a characteristic of weak boson localization.
}
\label{figure5} 
\end{center}
\end{figure}

Let us discuss the analysis of the weakly- and strongly-localized regions of the observed data using the conventional $\beta$-function, $\beta \equiv \mathrm{d}(\mathrm{ln}~g)/\mathrm{d}(\mathrm{ln}~L)$, where $L$ is the sample size and $g$ is the dimensionless conductance $[g=(\hbar/e^2)\sigma]$.
The sample size $L$ is a cutoff length due to inelastic scattering: $L^2=DT^{-p}$, where $D$ is a diffusion constant of electrons.
Here, we set $p=1$.
The $\beta$-function is described as
\begin{equation}
    \beta_{\mathrm{EXP.}} (g) = -\frac{2}{p} \frac{\mathrm{d}~\mathrm{ln}~g}{\mathrm{d}~\mathrm{ln}~T}.
\label{eq:beta_func}
\end{equation}
Using this equation, we have plotted the conductance data in the Fig.~\ref{figure5}. 
Vollhardt and W\"{o}lfle have reported a self-consistent theory from weak to strong localization~\cite{vollhardt1982scaling}.
According to their theory, for $g \ll 1$, the $\beta_{\mathrm{VW}}$-function is proportional to $\mathrm{ln}~g$.
For $g>1$, the $\beta_{\mathrm{VW}}$ is proportional to $-1/g$, and $\beta=0$ for $g \gg 1$.
We exhibit $\beta_{\mathrm{VW}}(g)$ function in Fig.~\ref{figure5} as a solid pink curve.
Surprisingly, we can see that all of the data fall on a single universal curve from weak to strong localized regions, as shown in Fig.~\ref{figure5}(a).
In particular, for samples 4-7, we observed a flow that rises perpendicularly from the universal curve in the weak localization region.
This vertical flow indicates the transition from the Fermi glass to the Bose glass in the weak localization region.
The vertical flow behavior is consistent with the results of the $\mathrm{ln}~T$ dependence of the resistance in Fig.~\ref{figure3}.
We have shown that the $\beta$-function scaling analysis can be applied to \CRO, such as copper oxides and Pb films.

\begin{figure}[t]
\begin{center}
\includegraphics[width=1\linewidth]{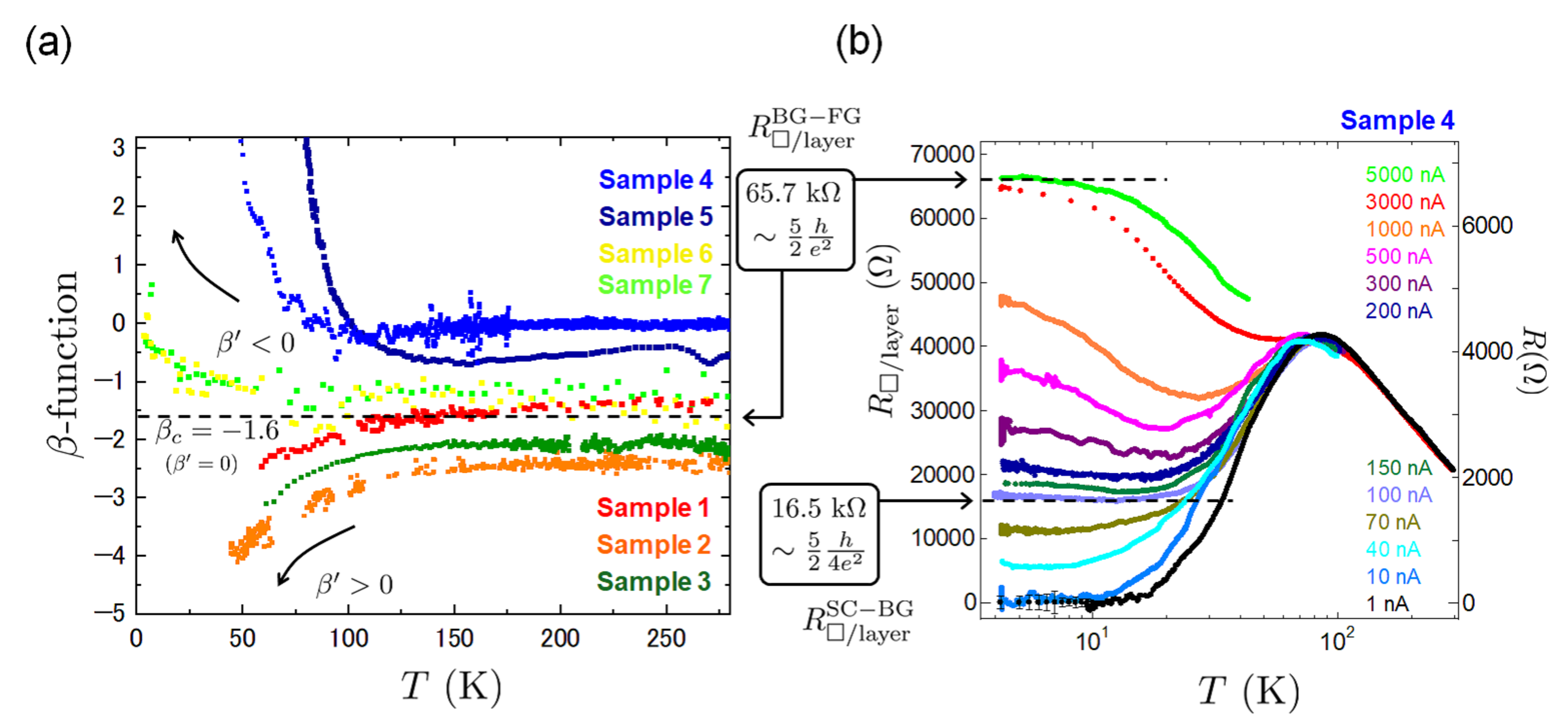}
\caption{
    \textbf{Observation of two quantum critical resistances.}
    (a) Temperature dependence of the $\beta$-function and the critical value $\beta_c$.
    As discussed in the main text, when the temperature derivative $\beta'$ of $\beta(g)$ is positive, it exhibits the behavior of the Fermi glass. 
    For $\beta' <0$, it becomes the Bose glass.
    This plot shows the critical value $\beta_{c}=-1.6$ between the Bose glass and Mott insulating phases.  
    The critical sheet resistance estimated from this value is $R_{\Box / \mathrm{layer}}^{\mathrm{BG-FG}}=65.7~\mathrm{k}\Omega \sim \frac{5}{2}\frac{h}{e^2}$.
    (b) Temperature dependence of the sheet resistance $R_{\Box / \mathrm{layer}}$ with applied bias current for sample 4.
    The critical sheet resistances of $R_{\Box / \mathrm{layer}}^{\mathrm{SC-BG}}=16.5~\mathrm{k}\Omega$ and $R_{\Box / \mathrm{layer}}^{\mathrm{BG-FG}}=65.7~\mathrm{k}\Omega$ were observed, respectively.
}
\label{figure6} 
\end{center}
\end{figure}

We discuss the Bose and Fermi glass transitions regarding the temperature dependence of the $\beta_{\mathrm{EXP.}}$-function.
The change in the flow of the $\beta_{\mathrm{EXP.}}$-function is expressed in the sign of the temperature derivative of the $\beta_{\mathrm{EXP.}}$-function.
The temperature derivative of the $\beta_{\mathrm{EXP.}}$-function is described as follows.
\begin{equation}
    \beta'_{\mathrm{EXP.}}(g)=-\frac{\mathrm{d}~\mathrm{ln}~|\beta_{\mathrm{EXP.}}(g)|}{\mathrm{d}~\mathrm{ln}~T}.
\label{eq:beta'}
\end{equation}
By substituting the equations $\sigma = \sigma_0 + \sigma_1 \mathrm{ln}~T $ and $\sigma = 1/\rho = 1/(\rho_0 + \rho_1 \mathrm{ln}~(1/T))~(\sigma_0,~\sigma_1,~\rho_0,~\rho_1>0)$, we obtain that the signs are different as follows.

\begin{equation}
    \beta'_{\mathrm{EXP.}}(g)=\frac{\sigma_1}{\sigma_0 + \sigma_1 \mathrm{ln}~T}>0,
\label{eq:beta'FG}
\end{equation}

\begin{equation}
    \beta'_{\mathrm{EXP.}}(g)=- \frac{\rho_1}{\rho_0 + \rho_1 \mathrm{ln}~(1/T)}<0.
\label{eq:beta'BG}
\end{equation}
When the temperature derivative of the $\beta_{\mathrm{EXP.}}$-function is positive, it indicates a Fermi glass; when it is negative, it signifies a Bose glass~\cite{takahashi2023bose}.
Figure~\ref{figure6} shows the result of plotting the $\beta_{\mathrm{EXP.}}$-function for temperature.
The sign of $\beta'_{\mathrm{EXP.}}$ allows us to distinguish between the Bose-Fermi glass transition in the weak localization region.
Samples 4-7 show the behavior of an upturn ($\beta'_{\mathrm{EXP.}}<0$), which can be identified as the Bose glass. 
On the other hand, samples 1-3 show the behavior of downturn ($\beta'_{\mathrm{EXP.}}>0$), which is the Fermi glass.
Furthermore, we determined the Bose glass and Fermi glass transition's quantum critical point $\beta_c=-1.6$, from Fig.~\ref{figure6}(a).
The quantum sheet resistance estimated from $g_c=-1.6$ is $R_{\Box / \mathrm{layer}}^{\mathrm{BG-FG}}=65.7~\mathrm{k}\Omega \sim\frac{5}{2}\frac{h}{e^2}$.
This value is 2.5($=5/2$) times larger than the universal resistance $h/e^2$ clarified by the Bose-glass-Fermi-glass transition~\cite{takahashi2023bose}.
Interestingly, we also observed a quantum resistance of 65.7~k$\Omega$ in the bias current-induced SIT~\cite{Nobukane_CRO} for sample 4, as shown in Fig.~\ref{figure6}(b).
Moreover, in Fig.~\ref{figure6}(b), we observed a quantum sheet resistance of $R_{\Box / \mathrm{layer}}^{\mathrm{SC-BG}}=16.8$ $\mathrm{k}\Omega \sim\frac{5}{2}\frac{h}{4e^2}$ at the superconductor-Bose glass transition, which is 2.5($=5/2$) times larger than the quantum resistance $h/4e^2$ value in 2D SIT.
The critical sheet resistance at the two quantum phase transition points in \CRO is 2.5 times larger and may be related to the contribution of strong electron correlation.
Fisher \textit{et al.}~\cite{fisher1990presence}, theoretically pointed out that the quantum resistance of $h/4e^2$ in the superconductor-Mott insulator transition is multiplied by the factor $8/\pi (\sim 2.55$).
Another example of a quantum resistance coefficient of 5/2 is the Pfaffian state in the fractional quantum Hall system~\cite{moore1991nonabelions}.

We have demonstrated that bosons appear at high temperatures above the $T_c^{\mathrm{onset}}$, even though the 2D superconducting thin films show finite resistance.
The existence of the Bose glass phase that appears between the superconducting and insulating phases may apply to recent nickel oxide superconductivity in addition to cuprates~\cite{takahashi2023bose} and ruthenates~\cite{Nobukane_CRO}.
Nickelates have been reported to have weak localization and strange metal phases near the superconducting phase~\cite{li2019superconductivity,zhang2024high}, and interest is growing in their relationship to the Bose glass phase.
In particular, superconductivity has been reported in thin films at ambient pressure~\cite{zhou2025ambient}, and the existence of the Bose glass phase may be important in understanding perovskite oxide superconductors.

\noindent
\subsection*{Conclusion}
We investigated the localization region of the \CRO~ in detail by performing a scaling analysis of the $\beta$-function. 
Making the film thinner suppresses the strong localization (Mott insulating phase), and the Bose glass and superconducting phases become the ground state. 
Interestingly, the weak localized bosons occur at the highest temperature, around 220~K.
We determined the quantum sheet resistances of $R_{\Box / \mathrm{layer}}^{\mathrm{BG-FG}}=\frac{5}{2}\frac{h}{e^2}$ and $R_{\Box / \mathrm{layer}}^{\mathrm{SC-BG}}=\frac{5}{2}\frac{h}{4e^2}$ at these two quantum critical points. 
The ground state of the \CRO~ can be understood by observing the change in electrical resistivity from room temperature to 120~K. 
In other words, as we suppress localization, Cooper pairs form. 
Thinning in ruthenates does not introduce disorder but contributes to controlling electronic correlations by relieving the distortion in RuO$_6$ octahedra.

\noindent
\subsection*{Methods}
To obtain nanoscale Ca$_2$RuO$_4$ single crystals, we synthesized Ca$_2$RuO$_4$ crystals with a solid phase reaction.
The method for growing single crystals is detailed in Ref.~\cite{Nobukane_CRO}.
The structure of Ca$_2$RuO$_4$ crystals was analyzed using X-ray diffraction (Rigaku Miniflex600) with Cu $K\alpha$ radiation.
The samples were dispersed in dichloroethane using sonication and then deposited onto a SiO$_2$ (300~nm) / Si substrate.
We could distinguish Ca$_2$RuO$_4$ and CaRuO$_3$ by observing the crystal shape using a scanning electron microscope.
We then fabricated gold electrodes using standard electron beam lithography methods.
The sample thickness was determined using scanning electron micrographs with a 70-degree tilt of the sample holder.

We also grew \CRO~ thin films on LaAlO$_3$ substrates using molecular beam epitaxy (MBE) methods~\cite{sakoda2021extraordinary}.
We performed X-ray diffraction to investigate the crystal structure of the thin film. 
The (006) peak corresponding to the $c$-axis was observed. The $c$-axis length was estimated from the interplanar spacing as $c=11.592$\r{A}.
This length is shorter than a nanoscale single crystal's $c$-axis length (12.278\r{A}).
Therefore, the distortion of the RuO$_6$ octahedra is thought to be more assertive in MBE thin films than in nanoscale single crystals.

We performed the electric transport measurements by the four-terminal method using a homemade $^3$He refrigerator. 
All leads were equipped with $RC$ filters ($R=1$~k$\Omega$ and $C=22$~nF).
In the DC measurements, a DC current source (6220, Keithley) supplied a bias current, and the voltage was measured with a nanovoltmeter (2182, Keithley).

\noindent
\subsection*{Acknowledgements}

We thank K. Yanagihara, Y. Ogasawara, K. Onodera, N. Matsunaga, K. Nomura, K. Inagaki, K. Nakatsugawa, H. Obuse, K. Ichimua, Y. Kunisada, K. Tanahashi, and T. Nomura for their experimental help and valuable discussions.
This work was supported by the Nippon Sheet Glass Foundation for Material Science and Engineering, the Yashima Environment Technology Foundation, the Inamori Foundation, the Iketani Science and Technology Foundation, and the Samco Science and Technology Foundation.

\subsection*{Author contributions}
H.N. designed the experiments.
K.H., K.I. M.H., M.S. and H.N. synthesized the high-quality samples and analysed the crystal structure.
K.H., K.I., M.H., M.S., K.T. and H.N. performed the transport measurements and analysed the data.
H.N. and S.T. contributed to the interpretation of the results.
H.N. drafted the manuscript.
All the authors read and approved the final manuscript.

\subsection*{Competing Interests}
The authors declare that they have no competing interests.

\bibliographystyle{naturemag}
\bibliography{Ruthenates_v2}

\newpage

\clearpage  

\renewcommand{\thesection}{S\arabic{section}}
\renewcommand{\theequation}{S.\arabic{equation}}
\renewcommand{\thefigure}{S\arabic{figure}}
\renewcommand{\thetable}{S\arabic{table}}

\setcounter{section}{0}
\setcounter{equation}{0}
\setcounter{figure}{0}
\setcounter{table}{0}

\begin{center}
\subsection*{Supplementary Information for `Bose glass in Ca$_2$RuO$_4$ nanofilms'}
\end{center}

\begin{figure}[h]
\begin{center}
\includegraphics[width=0.5\linewidth]{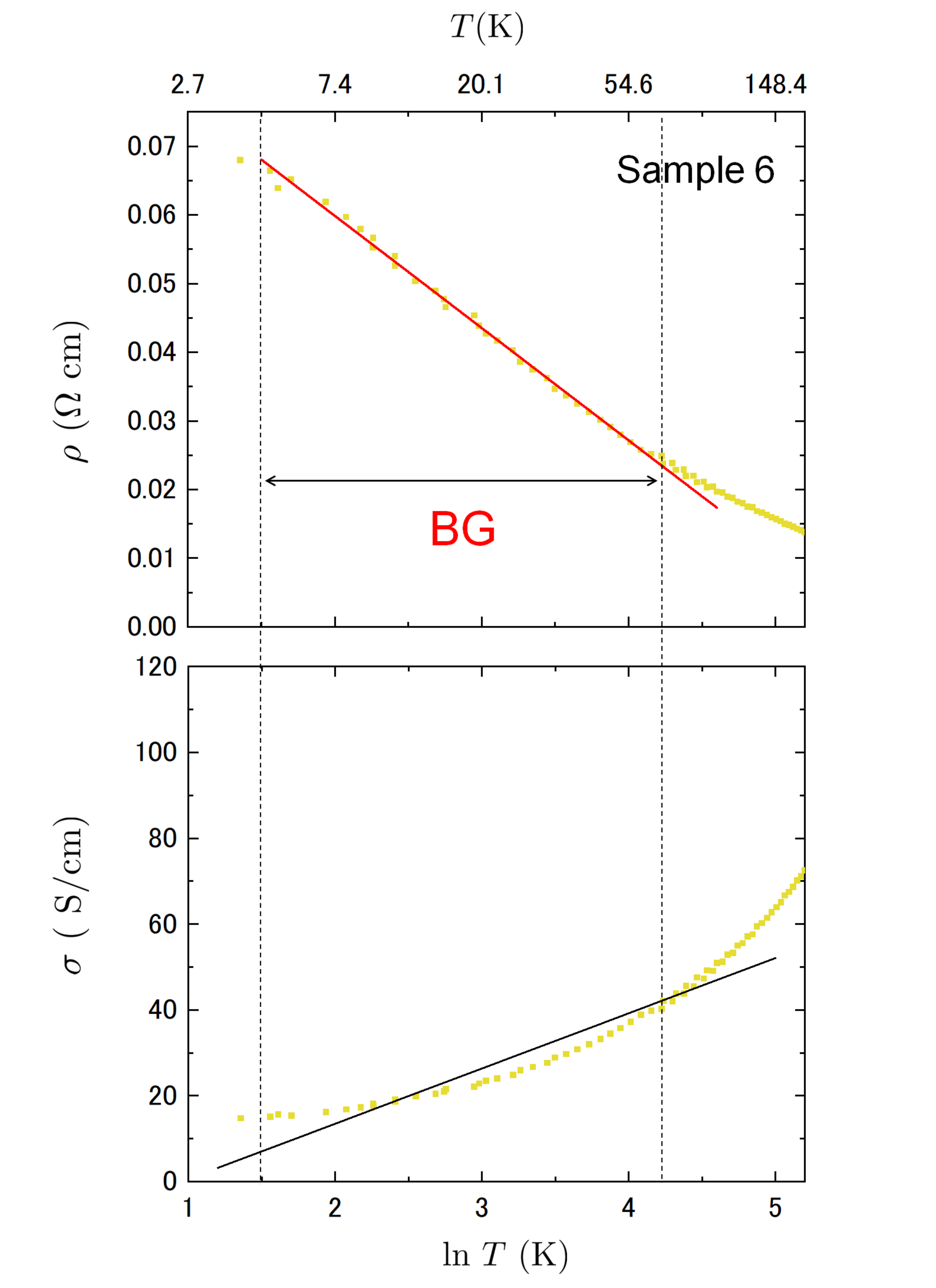}
 \caption{
   ln~$T$ dependence of resistivity and conductivity for MBE thin films (sample 6).
   }
\label{Supplfigure1}
\end{center}
\end{figure}

\newpage
\begin{figure}[h]
    \begin{center}
    \includegraphics[width=1\linewidth]{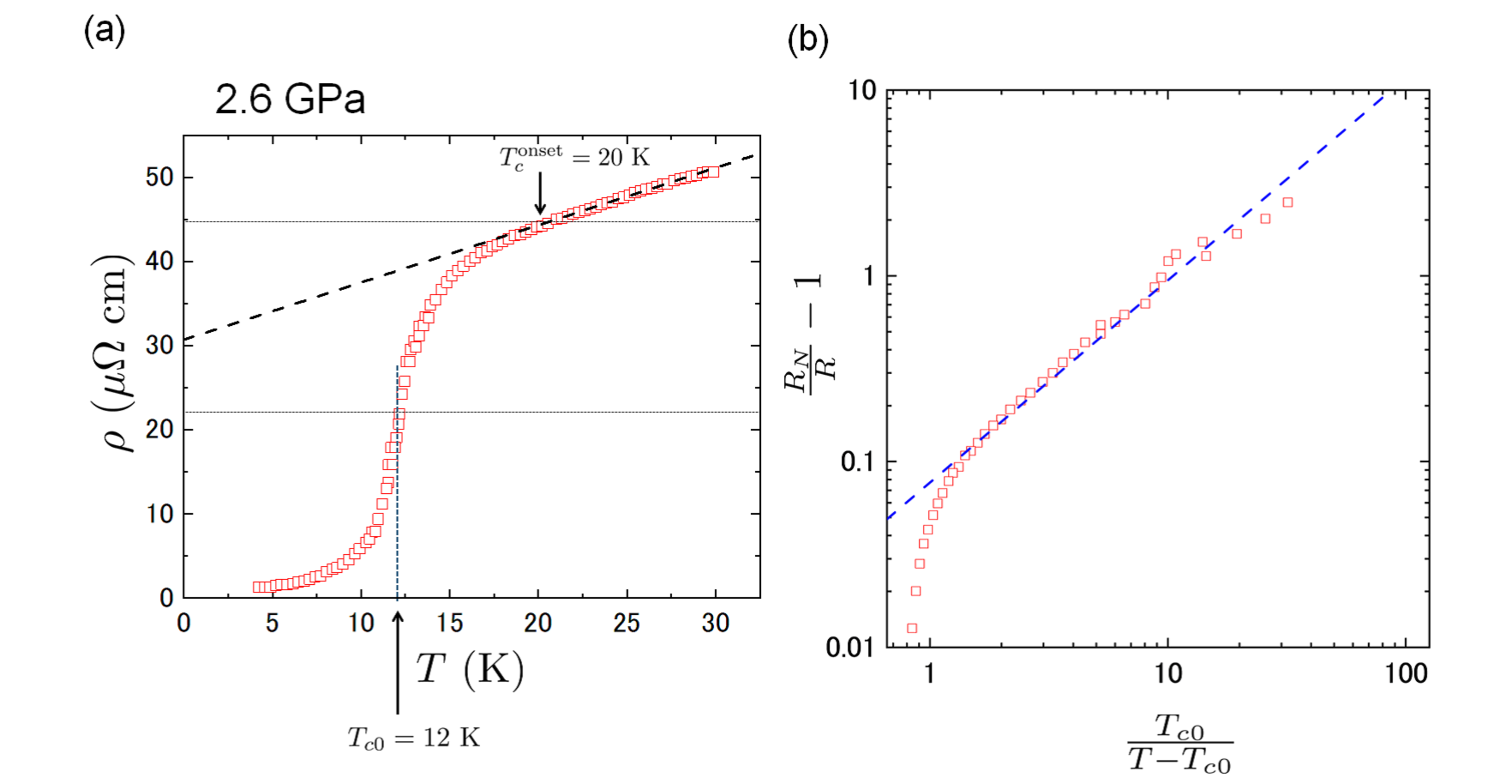}
     \caption{
      (a) Temperature dependence of resistivity in Fig.~3(c) of Ref.~\cite{nakamura2013electric}.
      (b) Result of superconducting fluctuation analysis.
    }
    \label{Supplfigure2}
    \end{center}
    \end{figure}

Figure \ref{Supplfigure2}(a) shows the temperature dependence of in-plane resistivity under 2.6~GPa in bulk \CRO~\cite{nakamura2013electric}.
Ref.~\cite{nakamura2013electric} concludes that this decrease in resistance is due to ferromagnetic metals. However, we are focusing on the possibility that this behavior is due to superconductivity. 
We performed a superconducting fluctuation analysis using the Aslamazov-Larkin (AL) model in two dimensions, $\frac{R_N}{R}-1=\frac{e^2}{16\hbar}R_N(\frac{T_{c0}}{T-T_{c0}})$, where we set $T_{c0}=12$~K and $R_{N}=44~\mu\Omega\cdot\mathrm{cm}$. 
The result fits the AL model as shown in Fig.~\ref{Supplfigure2}(b). 
The result suggests that this behavior is indicative of superconducting fluctuations.
It is also possible that \CRO~ coexists with superconductivity and ferromagnetism.

\end{document}